\def\conf{0}
\def\icalp{0}

\ifnum\icalp=0
  \documentclass[10pt]{article}
  \usepackage{fullpage}
\else
  \documentclass{llncs}
\fi

\usepackage{graphicx,amsfonts,amsmath,amssymb,epsfig,hyperref,color}
\usepackage{algorithm}
\usepackage{pdfpages}
\usepackage{enumitem}
\usepackage{multirow}
\usepackage{thm-restate}

\usepackage[noend]{algpseudocode}
\usepackage[english]{babel}
\usepackage[utf8]{inputenc}
\usepackage{amsmath}
\usepackage{graphicx}
\usepackage[colorinlistoftodos]{todonotes}

\usepackage{verbatim}
\usepackage{comment}
\usepackage{hyperref}

\usepackage{boxedminipage}
\usepackage{fullpage}
\usepackage{array}
\usepackage[normalem]{ulem}

\usepackage{hyperref}

\usepackage{caption}
\usepackage{subcaption}

\makeatother
\newcommand{\mnote}[1]{}

\renewcommand{\paragraph}[1]{{\protect\vspace{8pt}\noindent\sc{#1}}}

\newlength{\saveparindent}
\newlength{\saveparskip}
\newcommand{\BE}{\begin{enumerate}} \newcommand{\EE}{\end{enumerate}}
\newcommand{\BI}{\begin{itemize}} \newcommand{\EI}{\end{itemize}}
\newcommand{\BDes}{\begin{description}}\newcommand{\EDes}{\end{description}}
\newtheorem{alg}{Algorithm}
\newcommand{\BA}{\begin{alg}} \newcommand{\EA}{\end{alg}}
\newtheorem{thm}{Theorem}[section]            
\newcommand{\BT}{\begin{thm}} \newcommand{\ET}{\end{thm}}

\newtheorem{lem}[thm]{Lemma} 
\newcommand{\BL}{\begin{lem}} \newcommand{\EL}{\end{lem}}

\newtheorem{fact}{Fact}      
\newcommand{\BF}{\begin{fact}} \newcommand{\EF}{\end{fact}}
\newtheorem{clm}[thm]{Claim}
\newcommand{\BCM}{\begin{clm}} \newcommand{\ECM}{\end{clm}}

\newtheorem{techcor}[thm]{Corollary}
\newcommand{\BCo}{\begin{techcor}} \newcommand{\ECo}{\end{techcor}}

\newtheorem{cor}[thm]{Corollary}      
\newcommand{\BC}{\begin{cor}} \newcommand{\EC}{\end{cor}}
\newtheorem{prop}[thm]{Proposition}     
\newcommand{\BP}{\begin{prop}} \newcommand {\EP}{\end{prop}}
\newtheorem{conj} {Conjecture}      
\newcommand{\BCJ}{\begin{conj}} \newcommand{\ECJ}{\end{conj}}

\newtheorem{defn}{Definition}         
\newcommand{\BD}{\begin{defn}} \newcommand{\ED}{\end{defn}}

\def\FullBox{\hbox{\vrule width 8pt height 8pt depth 0pt}}
\newcommand{\qed}{\;\;\;\FullBox}
\newcommand{\ourqed}{\;\;\;\FullBox}
\newenvironment{ourproof}{\noindent{\bf Proof:~~}}{\(\ourqed\)}
\newcommand{\BPF}{\begin{ourproof}} \newcommand {\EPF}{\end{ourproof}}
\newenvironment{proofof}[1]{\noindent{\bf Proof of {#1}:~~}}{\(\qed\)}
\newcommand{\BPFOF}{\smallskip \begin{proofof}} \newcommand {\EPFOF}{\end{proofof}}
\newcommand{\qedsketch}{\;\;\;\Box}

\newenvironment{smallproof}{\noindent{\bf Proof:~~}}{\(\qedsketch\)}
\newcommand{\bpf}{\begin{smallproof}} \newcommand{\epf}{\end{smallproof}}

\newcommand{\BEQ}{\begin{equation}} \newcommand{\EEQ}{\end{equation}}
\newcommand{\BEQN}{\begin{eqnarray}}\newcommand{\EEQN}{\end{eqnarray}}

\newcommand{\eqdef}{\stackrel{\rm def}{=}}

\renewcommand{\Pr}{{\rm Pr}}

\newcommand{\Var}{{\mbox{\bf\rm Var}}}

\newcommand{\eps}{\epsilon}

\newcommand{\E}{{\rm E}}

\renewcommand{\O}{{\rm O}}

\renewcommand{\P}{{\rm P}}

\ifnum\conf=0
\fi

\ifnum\conf=1
\fi
\ifnum\conf=1
\subtitle{Extended abstract}
\else
\fi

  \title{Sublinear-Time Algorithms for Counting Star Subgraphs\\ with Applications to Join Selectivity Estimation} 
\ifnum\icalp=1
\author{Maryam Aliakbarpour\inst{1} Amartya Shankha Biswas\inst{2} Themistoklis Gouleakis\inst{3} John Peebles\inst{4} Ronitt Rubinfeld\inst{5} \and Anak Yodpinyanee\inst{6}
}
\authorrunning{Author Running?} 
\tocauthor{TOC Author?}
CSAIL, MIT, Cambridge MA 02139.
\email{ maryam.aliakbarpour@gmail.com},
\thanks{Research supported by NSF grants CCF-1217423, CCF-1065125, and CCF-1420692.}
\and
MIT, Cambridge MA 02139.
\email{ asbiswas@mit.edu},
\and
CSAIL, MIT, Cambridge MA 02139.
\email{ tgoule@mit.edu},
\thanks{Research supported by NSF grants CCF-1217423, CCF-1065125, and CCF-1420692.}
\and
CSAIL, MIT, Cambridge MA 02139.
\email{ johnpeeb@gmail.com},
\thanks{Research supported by NSF grants CCF-1217423, CCF-1065125, CCF-1420692, and CCF-1122374.}
\and
CSAIL, MIT, Cambridge MA 02139.
the Blavatnik School of Computer Science, Tel Aviv University.
\email{ ronitt@csail.mit.edu},
\thanks{Research supported by NSF grants CCF-1217423, CCF-1065125, CCF-1420692, and ISF grant 1536/14.}
\and
CSAIL, MIT, Cambridge MA 02139.
\email{ anak@csail.mit.edu},
\thanks{Research supported by NSF grants CCF-1217423, CCF-1065125, and CCF-1420692, and the DPST scholarship, Royal Thai Government.}
}
\else
\date{}
\author{
Maryam Aliakbarpour
\thanks{CSAIL, MIT, Cambridge MA 02139.
E-mail: {\tt  maryama@mit.edu}.
Research supported by NSF grants CCF-1217423, CCF-1065125, and CCF-1420692}
\and
Amartya Shankha Biswas
\thanks{MIT, Cambridge MA 02139.
E-mail: {\tt  asbiswas@mit.edu}.}
\and
Themistoklis Gouleakis
\thanks{CSAIL, MIT, Cambridge MA 02139.
E-mail: {\tt  tgoule@mit.edu}.
Research supported by NSF grants CCF-1217423, CCF-1065125, and CCF-1420692.}
\and
John Peebles
\thanks{CSAIL, MIT, Cambridge MA 02139.
E-mail: {\tt  johnpeeb@gmail.com}.
Research supported by NSF grants CCF-1217423, CCF-1065125, CCF-1420692, and CCF-1122374.}
\and
Ronitt Rubinfeld
\thanks{CSAIL, MIT, Cambridge MA 02139 and
the Blavatnik School of Computer Science, Tel Aviv University.
E-mail: {\tt  ronitt@csail.mit.edu}.
Research supported by NSF grants CCF-1217423, CCF-1065125, CCF-1420692, and ISF grant 1536/14.}
\and
Anak Yodpinyanee
\thanks{CSAIL, MIT, Cambridge MA 02139.
E-mail: {\tt  anak@csail.mit.edu}.
Research supported by NSF grants CCF-1217423, CCF-1065125, and CCF-1420692, and the DPST scholarship, Royal Thai Government.}
}
\fi

\begin{document}

\ifnum\icalp=0
\begin{titlepage}
\fi

\maketitle
%
%
%
\ifnum\icalp=0
%
%
%
%
%
\thispagestyle{empty}
\fi

\begin{abstract}
We study the problem of estimating the value of sums of the form $S_p \triangleq \sum \binom{x_i}{p}$ when one has the ability to sample $x_i \geq 0$ with probability proportional to its magnitude. When $p=2$, this problem is equivalent to estimating the selectivity of a self-join query in database systems when one can sample rows randomly. We also study the special case when $\{x_i\}$ is the degree sequence of a graph, which corresponds to counting the number of $p$-stars in a graph when one has the ability to sample edges randomly.

Our algorithm for a $(1 \pm \varepsilon)$-multiplicative approximation of $S_p$ has query and time complexities $\O(\frac{m \log \log n}{\epsilon^2 S_p^{1/p}})$. 
Here, $m=\sum x_i/2$ is the number of edges in the graph, or equivalently, half the number of records in the database table. Similarly, $n$ is the number of vertices in the graph and the number of unique values in the database table. We also provide tight lower bounds (up to polylogarithmic factors) in almost all cases, even when $\{x_i\}$ is a degree sequence and one is allowed to use the structure of the graph to try to get a better estimate. We are not aware of any prior lower bounds on the problem of join selectivity estimation.

For the graph problem, prior work which assumed the ability to sample only \emph{vertices} uniformly gave algorithms with matching lower bounds [Gonen, Ron, and Shavitt. \textit{SIAM J. Comput.}, 25 (2011), pp. 1365-1411]. With the ability to sample edges randomly, we show that one can achieve faster algorithms for approximating the number of star subgraphs, bypassing the lower bounds in this prior work. For example, in the regime where $S_p\leq n$, and $p=2$, our upper bound is $\tilde{O}(n/S_p^{1/2})$, in contrast to their $\Omega(n/S_p^{1/3})$ lower bound when no random edge queries are available.

In addition, we consider the problem of counting the number of directed paths of length two when the graph is directed. This problem is equivalent to estimating the selectivity of a join query between two distinct tables.
We prove that the general version of this problem cannot be solved in sublinear time. However, when the ratio between in-degree and out-degree is bounded---or equivalently, when the ratio between the number of occurrences of values in the two columns being joined is bounded---we give a sublinear time algorithm via a reduction to the undirected case.
\end{abstract}

\ifnum\icalp=0
\end{titlepage}
\fi


\section{Introduction}
We study the problem of approximately estimating $S_p \triangleq \sum_{i=1}^n {x_i \choose p}$ when one has the ability to sample $x_i \geq 0$ with probability proportional to its magnitude. To solve this problem we design \emph{sublinear-time algorithms}, which compute such an approximation while only looking at an extremely tiny fraction of the input, rather than having to scan the entire data set in order to determine this value.

We consider two primary motivations for this problem. The first is that in undirected graphs, if $x_i$ is the degree of vertex $i$ then $S_p$ counts the number of $p$-stars in the graph. Thus, estimating $S_p$ when one has the ability to sample $x_i$ with probability proportional to its magnitude corresponds to estimating the number of $p$-stars when one has the ability to sample vertices with probability proportional to their degrees (which is equivalent to having the ability to sample edges uniformly). This problem is an instance of the more general \emph{subgraph counting problem} in which one wishes to estimate the number of occurrences of a subgraph $H$ in a graph $G$. The subgraph counting problem has applications in many different fields, including the study of biological, internet and database systems.
For example, detecting and counting subgraphs in protein interaction networks is used to study molecular pathways and cellular processes across species \cite{scott2006efficient}.

The second application of interest is that the problem of estimating $S_2$ corresponds to estimating the selectivity of join and self-join operations in databases when one has the ability to sample rows of the tables uniformly. For example, note that if we set $x_i$ as the number of occurrences of value $i$ in the column being joined, then $S_2$ is precisely the number of records in the join of the table with itself on that column. When performing a query in a database, a program called a \emph{query optimizer} is used to determine the most efficient way of performing the database query. In order to make this determination, it is useful for the query optimizer to know basic statistics about the database and about the query being performed. For example, queries that return a very larger number of records are usually serviced most efficiently by doing simple linear scans over the data whereas queries that return a smaller number of records may be better serviced by using an index \cite{haas2009discovering}. As such, being able to estimate \emph{selectivity} (number of records returned compared to the maximum possible number) of a query can be useful information for a query optimizer to have. In the more general case of estimating the selectivity of a join between two different tables (which can be modeled with a directed graph), the query optimizer can use this information to decide on the most efficient order to execute a sequence of joins which is a common task.

In the ``typical'' regime in which we wish to estimate $S_2$ given that $n \leq S_2 \leq n^2$, our algorithm has a running time of $\tilde{O}(\sqrt{n})$ which is very small compared to than the total amount of data. Furthermore, in the case of selectivity estimation, this number can be much less than the number of distinct values in the column being joined on, which results in an even smaller number of queries than would be necessary if one were using an index to compute the selectivity.

We believe that our query-based framework can be realized in many systems.
One possible way to implement random edge queries is as follows: because edges normally take most of the space for storing graphs, an access to a random memory location where the adjacency list is stored, would readily give a random edge.
Random edge queries allow us to implement a source of {\em weighted
vertex samples}, where a vertex is output with probability proportional to its weight (magnitude).
Weighted sampling is used in \cite{motwani2007estimating,batu2009sublinear} to find sublinear algorithms for approximating the sum of $n$ numbers (allowing only uniform sampling, results in a linear lower bound). We later use this as a subroutine in our algorithm.

Throughout the rest of the paper, we will mostly use graph terminology when discussing this problem. However, we emphasize that all our results are fully general and apply to the problem of estimating $S_p$ even when one does not assume that the input is a graph.

\subsection{Our Contribution}

Prior theoretical work on this problem only considered the version of this problem on graphs and assumed the ability to sample vertices uniformly rather than edges. 
Specifically, prior studies of sublinear-time algorithms for graph problems usually consider the model where
the algorithm is allowed to query the adjacency list representation of the graph:
it may make \emph{neighbor queries} (by asking ``what is the $i^{\textrm{th}}$ neighbor of a vertex $v$'') and \emph{degree queries} (by asking ``what is the degree of vertex $v$'').

We propose a stronger model of sublinear-time algorithms for graph problems which allows random edge queries.
Next, for undirected graphs, we construct an algorithm which uses only degree queries and random edge queries. This algorithm and its analysis is discussed in Section~\ref{upper-undirected}.
For the problem of computing an approximation $\hat{S}_p$ satisfying $(1-\epsilon)S_p \leq \hat{S}_p \leq (1+\epsilon)S_p$, our algorithm has query and time complexities $\O(m \log \log n / \epsilon^2 S_p^{1/p})$. 
Although our algorithm is described in terms of graphs, it also applies to the more general case when one wants to estimate $S_p = \sum_i \binom{x_i}{p}$ without any assumptions about graph structure. Thus, it also applies to the problem of self-join selectivity estimation.

We then establish some relationships between $m$ and other parameters so that we may compare the performance of this algorithm
to a related work by Gonen et al.~more directly (\cite{gonen2011counting}).
We also provide lower bounds for our proposed model in Section~\ref{lower-undirected-section}, which are mostly tight up to polylogarithmic factors.
This comparison is given in Table \ref{undirected-results}. We emphasize that even though these lower bounds are stated for graphs, they also apply to the problem of self-join selectivity estimation.

To understand this table, first note that these algorithms require more samples when $S_p$ is small (i.e., stars are rare).
As $S_p$ increases, the complexity of each algorithm decreases until---at some point---the number of required samples drops to $\tilde{\O}(n^{1-1/p})$.
Our algorithm is able to obtain this better complexity of $\tilde{\O}(n^{1-1/p})$ for a larger range of values of $S_p$ than that of the algorithm given in \cite{gonen2011counting}. Specifically, our algorithm is more efficient 
for $S_p \leq n^{1+1/p}$, and has the same asymptotic bound for $S_p$ up to $n^p$. 
Once $S_p > n^p$, it is unknown whether the degree and random edge queries alone can provide the same query complexity. 
Nonetheless, if we have access to all three types of queries, we may combine the two algorithms to obtain the best of both cases as illustrated in the last column.

{\renewcommand{\arraystretch}{1.4}
\begin{table}[ht]
\centering
   \begin{tabular}{|c|c|c|c|} 
	\hline
	\multirow{3}{*}{range of $S_p$} & \multicolumn{3}{ c| }{permitted types of queries}  \\ \cline{2-4}
	& neighbor, degree  & degree, random edge & all types of queries \\
	& (\cite{gonen2011counting}) & (this paper) & (this paper) \\
	\hline
	$S_p \leq n$ & \multirow{2}{*}{$\widetilde{\Theta}\left(\frac{n}{S_p^{1/(p+1)}}\right)$} & $\widetilde{\Theta}\left(\frac{n}{S_p^{1/p}}\right)$ & $\widetilde{\Theta}\left(\frac{n}{S_p^{1/p}}\right)$ \\ \cline{1-1}\cline{3-4}
	$n < S_p \leq n^{1+1/p}$ & & \multirow{2}{*}{$\widetilde{\Theta}\left(n^{1-1/p}\right)$} & \multirow{2}{*}{$\widetilde{\Theta}\left(n^{1-1/p}\right)$} \\ \cline{1-2}
	$n^{1+1/p} < S_p \leq n^p$ & $\widetilde{\Theta}\left(n^{1-1/p}\right)$ & & \\ \hline
	$ n^p < S_p$ & $\widetilde{\Theta}\left(\frac{n^{p-1/p}}{S_p^{1-1/p}}\right)$ & $\Omega \left(\frac{n^{p-1/p}}{S_p^{1-1/p}}\right), \widetilde{\O}\left(n^{1-1/p}\right)$ & $\widetilde{\Theta}\left(\frac{n^{p-1/p}}{S_p^{1-1/p}}\right)$ \\ \hline
  \end{tabular}
\caption{Summary of the query and time complexities for counting $p$-stars on undirected graphs, given a different set of allowed queries. $\epsilon$ is assumed to be constant. Adjacent cells in the same column with the same contents have been merged.}\label{undirected-results}
\end{table}
}

We also consider a variant of the counting stars problem on directed graphs in Appendix~\ref{shortdir}.
If one only needs to count ``stars'' where all edges are either pointing into or away from the center, this is essentially still the undirected case.
We then consider counting directed paths of length two, and discover that allowing random edge queries does not provide an efficient algorithm in this case.
In particular, we show that any constant factor multiplicative approximation of $S_p$ requires $\Omega(n)$ queries even when all three types of queries are allowed.
However, when the ratio between the in-degree and the out-degree on every vertex is bounded, we solve this special case in sublinear time via a reduction to the undirected case where degree queries and random edge queries are allowed.

This variant of the counting stars problem can also be used for approximating join selectivity.
For a directed graph, we aim at estimating the quantity $\sum_{v \in V(G)} \deg^-(v)\cdot\deg^+(v)$.
On the other hand in the database context, we wish to compute the quantity $\sum_{i=1}^n x_i \cdot y_i$,
where $x_i$ and $y_i$ denote the number of occurrences of a label $i$ in the column we join on,
from the first and the second table, respectively.
Thus, applying simple changes in variables,
the algorithms from Appendix~\ref{shortdir} can be applied to the problem of estimating join selectivity as well.

\subsection{Our Approaches}

In order to approximate the number of stars in the undirected case, we convert the random edge queries into weighted vertex sampling,
where the probability of sampling a particular vertex is proportional to its degree.
We then construct an unbiased estimator that approximates the number of stars using the degree of the sampled vertex as a parameter.
The analysis of this part is roughly based on the variance bounding method used in \cite{alon1996space}, which aims to approximate the frequency moment in a streaming model.
The number of samples required by this algorithm depends on $S_p$, which is not known in advance.
Thus we create a guessed value of $S_p$ and iteratively update this parameter until it becomes accurate.

To demonstrate lower bounds in the undirected case, we construct new instances to prove
tight bounds for the case in which our model is more powerful than the traditional model.
In other cases, we provide a new proof to show that the ability to sample uniformly random edges does not necessarily allow better performance in counting stars.
Our proof is based on applying Yao's principle and providing an explicit construction of the hard instances,
which unifies multiple cases together and greatly simplifies the approach of \cite{gonen2011counting}.\footnote{One useful technique for giving lower bounds on sublinear time algorithms,
pioneered by \cite{blais2012property}, is to make use of a connection between
lower bounds in communication complexity and lower bounds on sublinear
time algorithms.   More specifically,  by giving a reduction from a communication
complexity problem to the problem we want to solve,  a lower bound
on the communication complexity problem yields a lower
bound on our problem.  In the past, this approach has led to simpler and cleaner sublinear time lower bounds for many problems.  Attempts at such an approach for reducing the set-disjointness problem in communication complexity  to our estimation problem on graphs run into the following difficulties:
First, as explained in \cite{DBLP:journals/eccc/Goldreich13a}, the straightforward reduction adds a logarithmic overhead, thereby weakening
the lower bound by the same factor.  Second, the reduction seems to work only in the case of sparse graphs. Although it is not clear if these difficulties are insurmountable, it seems that it will not give a  simpler argument than the approach that we present in this work.}

For the directed case, we prove the lower bound using a standard construction and Yao's principle.
As for the upper bound when the in-degree and out-degree ratios are bounded,
we use rejection sampling to adjust the sampling probabilities so that we may apply the unbiased estimator method from the undirected case. 

\subsection{Related Work}

Motivated by applications in a variety of areas, the subgraph detection and counting problem and its variations have been studied in many different works,
often under different terminology such as network motif counting or pathway querying (e.g., \cite{milo2002network,prvzulj2004modeling,wernicke2006efficient,scott2006efficient,shlomi2006qpath,grochow2007network,hormozdiari2007not,hales2008motifs,alon2008biomolecular}).
As this problem is NP-hard in general, many approaches have been developed to efficiently count subgraphs more efficiently
for certain families of subgraphs or input graphs (e.g., \cite{duke1995fast,alon1997finding,flum2004parameterized,kashtan2004efficient,alon2008biomolecular,alon2009balanced,vassilevska2009finding,williams2009finding,gonen2009approximating,kolountzakis2010efficient,ahn2012graph,amini2012counting,fomin2012faster}).
As for applications to database systems, the problem of approximating the size of the resulting table of a join query or a self-join query in various contexts has been studied in \cite{swami1994estimation,haas1996selectivity,alon1999tracking}. Selectivity and query optimization have been considered, e.g., in \cite{poosala1997selectivity,lee1999multi,getoor2001selectivity,markl2007consistent,haas2009discovering}.

Other works that study sublinear-time algorithms for counting stars are \cite{gonen2011counting} that aims to approximate the number of stars,
and \cite{feige2006sums,goldreich2008approximating} that aim to approximate the number of edges (or equivalently, the average degree).
Note that \cite{gonen2011counting} also shows impossibility results for approximating triangles and paths of length three in sublinear time when given uniform edge sampling, limiting us from studying more sophisticated subgraphs. Recent work by Eden, Levi and Ron (\cite{eden2015approximately}) and Seshadhri (\cite{seshadhri2015simpler}) provide sublinear time algorithms to approximate the number of triangles in a graph. However, their model uses adjacency matrix queries and neighbor queries.
The problem of counting subgraphs has also been studied in the streaming model (e.g., \cite{bar2002reductions,buriol2006counting,becchetti2008efficient,manjunath2011approximate,kane2012counting}).
There is also a body of work on sublinear-time algorithms for approximating various graph parameters (e.g., \cite{parnas2007approximating,nguyen2008constant,yoshida2009improved,hassidim2009local,onak2012near}).

Abstracting away the graphical context of counting stars, we may view our problem as finding a parameter of a distribution:
edge or vertex sampling can be treated as sampling according to some distribution. In vertex sampling, we have a uniform distribution and in edge sampling, the probabilities are proportional to the degree. The number of stars can be written as a function of the degrees.
Aside from our work, there are a number of other studies that make use of combined query types for estimating a parameter of a distribution.
Weighted and uniform sampling are considered in \cite{motwani2007estimating,batu2009sublinear}.
Their algorithms may be adapted to approximate the number of edges in the context of approximating graph parameters when given weighted vertex sampling, which we will later use in this paper.
A closely related problem in the context of distributions, is the task of approximating frequency moments, mainly studied in the streaming model (e.g., \cite{alon1996space,coppersmith2004improved,indyk2005optimal,bhuvanagiri2006simpler}).
On the other hand, the combination of weighted sampling and probability distribution queries is also considered (e.g., \cite{canonne2014testing}).

\section{Preliminaries} \label{prelim}

In this paper, we construct algorithms to approximate the number of stars in a graph under different types of query access to the input graph.
As we focus on the case of simple undirected graphs, we explain this model here and defer the description for the directed case to Appendix~\ref{shortdir}.

\subsection{Graph Specification}

Let $G=(V,E)$ be the input graph, assumed to be simple and undirected.
Let $n$ and $m$ denote the number of vertices and edges, respectively.
The value $n$ is known to the algorithm.
Each vertex $v\in V$ is associated with a unique ID from $[n]\eqdef\{1,\ldots,n\}$.
Let $\deg(v)$ denote the degree of $v$.

Let $p\geq 2$ be a constant integer.
A \emph{$p$-star} is a subgraph of size $p+1$,
where one vertex, called the \emph{center}, is adjacent to the other $p$ vertices.
For example, a $2$-star is an undirected path of length 2.
Note that a vertex may be a center for many stars, and a set of $p+1$ vertices may form multiple stars.
Let $S_p$ denote the number of occurrences of distinct stars in the graph.

Our goal is to construct a randomized algorithm that outputs a value that is within a $(1\pm\epsilon)$-multiplicative factor of the actual number of stars $S_p$.
More specifically, given a parameter $\epsilon > 0$, the algorithm must give an approximated value $\widehat{S}_p$ satisfying the inequality $(1-\epsilon)S_p \leq \widehat{S}_p \leq (1+\epsilon)S_p$
with success probability at least $2/3$.

\subsection{Query Access} \label{sampling}

The algorithm may access the input graph by querying the \emph{graph oracle}, which answers for the following types of queries.
First, the \emph{neighbor queries}: given a vertex $v\in V$ and an index $1 \leq i < n$, the $i^{\rm th}$ neighbor of $v$ is returned if $i \leq \deg (v)$; otherwise, $\bot$ is returned.
Second, the \emph{degree queries}: given a vertex $v \in V$, its degree $\deg(v)$ is returned.
Lastly, the \emph{random edge queries}: a uniformly random edge $\{u,v\} \in E$ is returned.
The \emph{query complexity} of an algorithm is the total number of queries of any type that the algorithm makes throughout the process of computing its answer.

Combining these queries, we may implement various useful sampling processes.
We may perform a \emph{uniform edge sampling} using a random edge query, and a \emph{uniform vertex sampling} by simply picking a random index from $[n]$.
We may also perform a \emph{weighted vertex sampling} where each vertex is obtained with probability proportional to its degree as follows:
uniformly sample a random edge, then randomly choose one of the endpoints with probability $1/2$ each.
Since any vertex $v$ is incident with $\deg(v)$ edges, then the probability that $v$ is chosen is exactly $\deg(v)/2m$, as desired.

\subsection{Queries in the Database Model}
Now we explain how the above queries in our graph model have direct interpretations in the database model.
Consider the column we wish to join on.
For each valid label $i$, let $x_i$ be the number of rows containing this label.
We assume the ability to sample rows uniformly at random.
This gives us a label $i$ with probability proportional to $x_i$, which is a weighted sample from the distribution of labels.
We also assume that we can also quickly compute the number of other rows sharing the same label with a given row
(analogous to making a degree query). For example, this could be done quickly using an index on the column. Note that if one has an index that is augmented with appropriate information, one can compute the selectivity of a self-join query exactly in time roughly $O(k \log n)$ where $k$ is the number of distinct elements in the column. However, our methods can give runtimes that are asymptotically much smaller than this.

\section{Upper Bounds for Counting Stars in Undirected Graphs} \label{upper-undirected}

In this section we establish an algorithm for approximating the number of stars, $S_p$, of an undirected input graph.
We focus on the case where only degree queries and random edge queries are allowed.
This illustrates that even without utilizing the underlying structure of the input graph,
we are still able to construct a sublinear approximation algorithm that outperforms other algorithms under the traditional model in certain cases.

\subsection{Unbiased Estimator Subroutine}  \label{unbiased-sec}

Our algorithm uses weighted vertex sampling to find stars.
Intuitively, the number of samples required by the algorithms should be larger when stars are rare because it takes more queries to find them.
While the query complexity of the algorithm depends on the actual value of $S_p$, our algorithm does not know this value in advance.
In order to overcome this issue, we devise a subroutine which---given a guess $\widetilde{S}_p$ for the value of $S_p$---will give a $(1\pm \epsilon)$ approximation of $S_p$ if $\widetilde{S}_p$ is close enough to $S_p$ or tell us that $\widetilde{S}_p$ is much larger than $S_p$. Then, we start with the maximum possible value of $S_p$ and guess multiplicatively smaller and smaller values for it until we find one that is close enough to $S_p$, so that our subroutine is able to correctly output a $(1 \pm \epsilon)$ approximation.

Our subroutine works by computing the average value of an unbiased estimator to $S_p$ after drawing enough weighted vertex samples.
To construct the unbiased estimator, notice first that the number of $p$-stars centered at a vertex $v$ is ${\deg(v) \choose p}$.\footnote{For our counting purpose, if $x < y$ then we define ${x \choose y} = 0$.}
Thus, $S_p = \sum_{v\in V} {\deg(v) \choose p}$.

Next, we define the unbiased estimator and give the corresponding algorithm.
First, let $X$ be the random variable representing the degree of a random vertex obtained through weighted vertex sampling, as explained in Section~\ref{sampling}.
Recall that a vertex $v$ is sampled with probability $\deg(v)/2m$.
We define the random variable $Y = \frac{2m}{X} {X \choose p}$ so that $Y$ is an unbiased estimator for $S_p$; that is,
\[E[Y]=\sum_{v\in V} \frac{\deg(v)}{2m}\left(\frac{2m}{\deg(v)} {\deg(v) \choose p}\right) = \sum_{v\in V} {\deg(v) \choose p} = S_p.\]

\begin{algorithm}
\label{alg:alg1}
\caption{Subroutine for Computing $S_p$ given $\widetilde{S}_p$ with success probability 2/3}\label{unbiased-algo}
\begin{algorithmic}[1]
\Procedure{Unbiased-Estimate}{$\widetilde{S}_p, \epsilon$}
	\State{$k \gets 36m\,/\,p \epsilon^2\widetilde{S}_p^{1/p}$}
	\For {$i = 1$ to $k$}
        \State{$v \gets $ weighted sampled vertex obtained from a random edge query}
    	  \State{$d \gets \deg(v)$ obtained from a degree query}
    	  \State{$Y_i \gets \frac{2m}{d} {d \choose p}$}
	\EndFor
    \State{ $\bar{Y} \gets \frac{1}{k} \sum_{i=1}^{k} Y_i$}
    \State{ \textbf{return} $\bar{Y} $}
\EndProcedure
\end{algorithmic}
\end{algorithm}

Clearly, the output $\bar{Y}$ of Algorithm~\ref{unbiased-algo} satisfies $E[\bar{Y}]=S_p$.
We claim that the number of samples $k$ in Algorithm \ref{unbiased-algo} is sufficient to provide two desired properties:
the algorithm returns an $(1\pm\epsilon)$-approximation of $S_p$ if $\widetilde{S}_p$ is in the correct range;
or, if $\widetilde{S}_p$ is too large, the anomaly will be evident as the output $\bar{Y}$ will be much smaller than $\widetilde{S}_p$.
In particular, we may distinguish between these two cases by comparing $\bar{Y}$ against $(1-\eps)\tilde{S}_p$, as specified through the following lemma. 

\begin{restatable}{lem}{gwc} \label{good-when-crude} \label{found-when-fault}
For $0 < \eps \leq 1/2$, with probability at least 2/3:
\begin{enumerate}[noitemsep,nolistsep]
\item If $\frac{1}{2}S_p \leq \widetilde{S}_p \leq 6S_p$, then Algorithm \ref{unbiased-algo} outputs $\bar{Y}$ such that $(1-\epsilon)S_p \leq \bar{Y} \leq (1+\epsilon)S_p$;\\
moreover, if $S_p < \tilde{S}_p$ then $\bar{Y} \geq (1-\epsilon)\widetilde{S}_p$.
\item If $\widetilde{S}_p > 6S_p$, then Algorithm \ref{unbiased-algo} outputs $\bar{Y}$ such that $\bar{Y} < \frac{1}{2}\widetilde{S}_p \leq (1-\epsilon)\widetilde{S}_p$.
\end{enumerate}
\end{restatable}

The first item of Lemma~\ref{good-when-crude} can be proved by bounding the variance of $Y$ using various Chebyshev's Inequality and identities of binomial coefficients,
while the second item is a simple application of Markov's Inequality.
Detailed proofs for these statements can be found in Appendix~\ref{upunpr}.

\subsection{Full Algorithm}

Our full algorithm proceeds by first setting $\widetilde{S}_p$ to $n{{n-1} \choose p}$, the maximum possible value of $S_p$ given by the complete graph.
We then use Algorithm \ref{unbiased-algo} to check if $\widetilde{S}_p > 6S_p$; if this is the case, we reduce $\widetilde{S}_p$ then proceed to the next iteration.
Otherwise, Algorithm \ref{unbiased-algo} should already give an $(1\pm\epsilon)$-approximation to $S_p$ (with constant probability).
We note that if $\epsilon > 1/2$, we may replace it with $1/2$ without increasing the asymptotic complexity.

Since the process above may take up to $\O(\log n)$ iterations,
we must amplify the success probability of Algorithm \ref{unbiased-algo} so that the overall success probability is still at least $2/3$.
To do so, we simply make $\ell = \O(\log \log n)$ multiple calls to Algorithm \ref{unbiased-algo} then take the median of the returned values.
Our full algorithm can be described as Algorithm \ref{full-algo} below.

\begin{algorithm}
\caption{Algorithm for Approximating $S_p$}\label{full-algo}
\begin{algorithmic}[1]
\Procedure{Count-Stars}{$\epsilon$}
	\State{$\widetilde{S}_p \gets n{{n-1} \choose p}, \; \ell \gets 40(\log p + \log \log n)$}
	\Loop
	\For {$i = 1$ to $\ell$}
        \State{$Z_i \gets \textsc{Unbiased-Estimate}(\widetilde{S}_p,\epsilon)$}
    	\EndFor
    	\State{$Z \gets \textrm{median}\{Z_1,\cdots,Z_\ell\}$}
        \If {$Z \geq (1-\epsilon)\widetilde{S}_p$}
        	\State {$\hat{S}_p \gets Z$}
            \State {\textbf{return} $\hat{S}_p$}
        \EndIf
	\State{$\widetilde{S}_p \gets \widetilde{S}_p/2$}
  \EndLoop
\EndProcedure
\end{algorithmic}
\end{algorithm}

\BT \label{upper-m}
Algorithm \ref{full-algo} outputs $\hat{S}_p$ such that $(1-\epsilon)S_p \leq \hat{S}_p \leq (1+\epsilon)S_p$ with probability at least $2/3$. The query complexity of Algorithm \ref{full-algo} is $\O\left(\frac{m \log n \log \log n}{\epsilon^2 S_p^{1/p}}\right)$.
\ET
\BPF
If we assume that the events from Lemma \ref{good-when-crude} hold,
then the algorithm will take at most $\lceil\log \left(n{{n-1} \choose p}\right)\rceil \leq (p+1) \log n$ iterations.
By choosing $\ell = 40(\log p + \log \log n)$, Chernoff bound (Theorem \ref{chernoff}) implies that excepted for probability $1/3(p+1) \log n$,
more than half of the return values of Algorithm \ref{unbiased-algo} satisfy the desired property, and so does the median $Z$.
By the union bound, the total failure probability is at most 1/3.

Now it is safe to assume that the events from the two lemmas hold.
In case $\widetilde{S}_p > 6S_p$, our algorithm will detect this event because $Z \leq (1-\epsilon)\widetilde{S}_p$,
implying that we never stop and return an inaccurate approximation.
On the other hand, if $\widetilde{S}_p < S_p$, our algorithm computes $Z \geq  (1-\epsilon)\widetilde{S}_p$ and must terminate.
Since we only halve $\widetilde{S}_p$ on each iteration, when $\widetilde{S}_p < S_p$ first occurs, we have $\widetilde{S}_p \geq \frac{1}{2}S_p$.
As a result, our algorithm must terminate with the desired approximation before the value $\widetilde{S}_p$ is halved again.
Thus, Algorithm \ref{full-algo} returns $\hat{S}_p$ satisfying $(1-\epsilon)S_p \leq \hat{S}_p \leq (1+\epsilon)S_p$ with probability at least $2/3$, as desired.

Recall that the number of samples required by Algorithm \ref{unbiased-algo} may only increase when $\widetilde{S}_p$ decreases.
Thus we may use the number of samples in the last round of Algorithm \ref{full-algo}, where $\widetilde{S}_p = \Theta(S_p)$, as the upper bound for each previous iteration.
Therefore, each of the $\O(\log n)$ iterations takes $\O(m \log \log n \,/\, \epsilon^2 S_p^{1/p})$ samples, achieving the claimed query complexity.
\EPF

\subsection{Removing the Dependence on $m$}

As described above, Algorithm \ref{unbiased-algo} picks the value $k$ and defines the unbiased estimator based on $m$, the number of edges.
Nonetheless, it is possible to remove this assumption of having prior knowledge of $m$ by instead computing its approximation.
Furthermore, we will bound $m$ in terms of $n$ and $S_p$, so that we can also relate the performance of our algorithm to previous studies on this problem such as \cite{gonen2011counting}, as done in Table \ref{undirected-results}.

\subsubsection{Approximating $m$} \label{approx-m}

We briefly discuss how to apply our algorithm when $m$ is unknown by first computing an approximation of $m$.
Using weighted vertex sampling, we may simulate the algorithm from \cite{motwani2007estimating} or \cite{batu2009sublinear} that computes an $(1\pm\epsilon)$-approximation to the sum of degrees using $\tilde{\O}(\sqrt{n})$ weighted samples.
More specifically, we cite the following theorem:
\BT (\cite{motwani2007estimating})
Let $x_1, \ldots, x_n$ be $n$ variables,
and define a distribution $\mathcal{D}$ that returns $(i, x_i)$ with probability $x_i / \sum_{i=1}^{n} x_j$.
There exists an algorithm that
computes a $(1\pm\epsilon)$-approximation of $S = \sum_{i=1}^{n} x_i$ using $\tilde\O(\sqrt{n})$ samples from $\mathcal{D}$.
\ET
Thus, we simulate the sampling process from $\mathcal{D}$ by drawing a weighted vertex sample $v$, querying its degree, and feeding $(v, \deg(v))$ to this algorithm.
We will need to decrease $\epsilon$ used in this algorithm and our algorithm by a constant factor to account for the additional error.
Below we show that our complexities are at least $\tilde{\O}(n^{1-1/p})$ which is already $\tilde{\O}(\sqrt{n})$ for $p = 2$, and thus this extra step does not affect our algorithm's performance asymptotically.

\subsubsection{Comparing $m$ to $n$ and $S_p$}

For comparison of performances, we will now show some bounds relating $m$ to $n$ and $S_p$.
Notice that the function ${\deg(v) \choose p}$ is convex with respect to $\deg(v)$.\footnote{We may use the binomial coefficients ${x \choose y}$ for non-integral value $x$ in the inequalities.
These can be interpreted through alternative formulations of binomial coefficients using falling factorials or analytic functions.}
Then by applying Jensen's inequality (Theorem \ref{jensen}) to this function, we obtain\[S_p = \sum_{v\in V} {\deg(v) \choose p} \geq n {{\sum_{v\in V} \deg(v) / n } \choose p} = n{2m/n\choose p}.\]

First, let us consider the case where the stars are very rare, namely when $S_p \leq n$.
The inequality above implies that $m \leq np/2$. Substituting this formula back into the bound from Theorem \ref{upper-m} yields the query complexity $\tilde{\O}(n\,/\,{\epsilon^2 S_p^{1/p}})$.

Now we consider the remaining case where $S_p > n$.
If $m < np/2 = \O(n)$, then the query complexity from Theorem \ref{upper-m} becomes $\tilde{\O}(n^{1-1/p}\,/\,{\epsilon^2})$.
Otherwise we have $2m/n \geq p$, which allows us to apply the following bound on our binomial coefficient:
\[S_p  \geq n{2m/n\choose p} \geq n\left(\frac{2m}{np}\right)^p.\]
This inequality implies that $m \leq pn^{1-1/p}S_p^{1/p}/2$, also yielding the query complexity $\tilde{\O}(n^{1-1/p}\,/\,{\epsilon^2})$.

Compared to \cite{gonen2011counting}, our algorithm achieves a better query complexity when $S_p \leq n^{1+1/p}$,
where the rare stars are more likely to be found via edge sampling rather than uniform vertex sampling or traversing the graph.
Our algorithm also performs no worse than their algorithm does for any $S_p$ as large as $n^p$.
Moreover, due to the simplicity of our algorithm, the dependence on $\epsilon$ of our query complexity is only $1/\epsilon^2$ for any value of $S_p$,
while that of their algorithm is as large as $1/\epsilon^{10}$ in certain cases.
This dependence on $\epsilon$ may be of interest to some applications, especially when stars are rare whilst an accurate approximation of $S_p$ is crucial.

\subsection{Allowing Neighbor Queries}

We now briefly discuss how we may improve our algorithm when neighbor queries are allowed (in addition to degree queries and random edge queries).
For the case when $S_p > n^p$, it is unknown whether our algorithm alone achieves better performance than \cite{gonen2011counting} (see table \ref{undirected-results}).
However, their algorithm has the same basic framework as ours, namely that it also starts by setting $\widetilde{S}_p$ to the maximum possible number of stars,
then iteratively halves this value until it is in the correct range, allowing the subroutine to correctly compute a $(1\pm\epsilon)$-approximation of $S_p$.
As a result, we may achieve the same performance as them in this regime by simply letting Algorithm \ref{full-algo} call the subroutine from \cite{gonen2011counting} when $S_p \geq n^p$.
We will later show tight lower bounds (up to polylogarithmic factors)
to the case where all three types of queries are allowed,
which is a stronger model than the one previously studied in their work.
\section{Lower Bounds for Counting Stars in Undirected Graphs} \label{lower-undirected-section}

In this section, we establish the lower bounds summarized in the last two columns of Table \ref{undirected-results}.
 We give lower bounds that apply even when the algorithm is permitted to sample random edges.
Our first lower bound is proved in Section \ref{smallcase};
While this is the simplest case, it provides useful intuition for the proofs of subsequent bounds.
In order to overcome the new obstacle of powerful queries in our model, for larger values of $S_p$
we create an explicit scheme for constructing families of graphs that are hard to distinguish by any algorithm even when these queries are present.
Using this construction scheme, our approach obtains the bounds for all remaining ranges for $S_p$ as special cases of a more general bound,
and the general bound is proved via the straightforward application of Yao's principle and a coupling argument.
Our lower bounds are tight (up to polylogarithmic factors) for all cases except for the bottom middle cell in Table \ref{undirected-results}.

\subsection{Lower Bound for $S_p \leq n$} \label{smallcase}

\BT \label{thm-small-bound}
For any constant $p \geq 2$, any (randomized) algorithm for approximating $S_p$ to a multiplicative factor via neighbor queries, degree queries and random edge queries with probability of success at least $2/3$ requires $\Omega(n/S_p^{1/p})$ total number of queries for any $S_p \leq n^p$.
\ET

\BPF
We now construct two families of graphs, namely $\mathcal{F}_1$ and $\mathcal{F}_2$, such that any $G_1$ and $G_2$ drawn from each respective family satisfy $S_p(G_1)=0$ and $S_p(G_2)=\Theta(s)$ for some parameter $s > (p+1)^p = O(1)$.
We construct $G_1$ as follows:
for a subset $S \subseteq V$ of size $\lceil s^{1/p} \rceil +1$,
we create a union of a $(p-1)$-regular graph on $S$ and a $(p-1)$-regular graph on $V \setminus S$, and add the resulting graph $G_1$ to $\mathcal{F}_1$. To construct all graphs in $\mathcal{F}_1$, we repeat this process for every subset $S$ of size $\lceil s^{1/p} \rceil +1$.
$\mathcal{F}_2$ is constructed a little differently:
rather than using a $(p-1)$-regular graph on $S$, we use a star of size $\lceil s^{1/p} \rceil$ on this set instead.
We add a union between a star on $S$ and a $(p-1)$-regular graph on $V \setminus S$ of any possible combination to $\mathcal{F}_2$.

By construction, every $G_1\in\mathcal{F}_1$ contains no $p$-stars, whereas every $G_2\in\mathcal{F}_2$ has ${{\O(s^{1/p})} \choose {p} } = \Theta(s)$ $p$-stars.
For any algorithm to distinguish between $\mathcal{F}_1$ and $\mathcal{F}_2$, when given a graph $G_2 \in \mathcal{F}_2$,
it must be able to detect some vertex in $S$ with probability at least $2/3$.
Otherwise, if we randomly generate a small induced subgraph according to the uniform distribution in $\mathcal{F}_2$ conditional on not having any vertex or edge in $S$, the distribution would be identical to the uniform in $\mathcal{F}_1$.
Furthermore, notice that $S$ cannot be reached via traversal using neighbor queries as it is disconnected from $V\setminus S$.
The probability of sampling such vertex or edge from each query is $\O(s^{1/p}/n)$.
Thus, $\Omega(n/s^{1/p})$ samples are required to achieve a constant factor approximation with probability 2/3.
\EPF

\subsection{Overview of the Lower Bound Proof for $S_p > n$} \label{largeboundsec}

Since graphs with large $S_p$ contain many edges, we must modify our approach above to allow graphs from the first family to contain stars.
We construct two families of graphs $\mathcal{F}_1$ and $\mathcal{F}_2$ such that the number of stars of graphs from these families differ by some multiplicative factor $c>1$;
any algorithm aiming to approximate $S_p$ within a multiplicative factor of $\sqrt{c}$ must distinguish between these two families with probability at least $2/3$.
We create \emph{representations} of graphs that explicitly specify their adjacency list structure.
Each $G_1 \in \mathcal{F}_1$ contains $n_1$ vertices of degree $d_1$, while the remaining $n_2 = n-n_1$ vertices are isolated.
For each $G_2\in \mathcal{F}_2$, we modify our representation from $\mathcal{F}_1$ by connecting each of the remaining $n_2$ vertices to $d_2 \gg d_1$ neighbors, so that these vertices contribute sufficient stars to establish the desired difference in $S_p$.
We hide these additional edges in carefully chosen random locations while ensuring minimal disturbance to the original graph representation; our representations are still so similar that any algorithm may not detect them without making sufficiently many queries.
Moreover, we define a coupling for answering random edge queries so that the same edges are likely to be returned regardless of the underlying graph. 

While the proof of \cite{gonen2011counting} also uses similar families of graphs, our proof analysis greatly deviates from their proof as follows.
Firstly, we apply Yao's principle which allows us to prove the lower bounds on randomized algorithms
by instead showing the lower bound on deterministic algorithms on our carefully chosen distribution of input instances.\footnote{See e.g., \cite{motwani2010randomized} for more information on Yao's principle.}
Secondly, rather than constructing two families of graphs via random processes,
we construct our graphs with adjacency list representations explicitly, satisfying the above conditions for each lower bound we aim to prove.
This allows us to avoid the difficulties in \cite{gonen2011counting} regarding the generation of potential multiple edges and self-loops in the input instances.
Thirdly, we define the distribution of our instances based on the permutation of the representations of these two graphs, and the location we place the edges in $G_2$ that are absent in $G_1$.
We also apply the coupling argument, so that the distribution of these permutations we apply on these graphs, as well as the answers to random edge queries, are as similar as possible.
As long as the small difference between these graphs is not discovered, the interaction between the algorithm and our oracle must be exactly the same.
We show that with probability $1 - o(1)$, the algorithm and our oracle behave in exactly the same way whether the input instance corresponds to $G_1$ or $G_2$.
Simplifying the arguments from \cite{gonen2011counting}, we completely bypass the algorithm's ability to make use of graph structures.
Our proof only requires some conditions on the parameters $n_1, d_1, n_2, d_2$;
this allows us to show the lower bounds for multiple ranges of $S_p$ simply by choosing appropriate parameters.

We provide the full details in Section ~\ref{lowerproof}.
The main results of our constructions are given as the following theorems.
We note that lower bounds apply when only subsets of these three types of queries are provided. This concludes all of our lower bounds in Table \ref{undirected-results}.

\BT \label{thm-medium-bound}
For any constant $p \geq 2$, any (randomized) algorithm for approximating $S_p$ to a multiplicative factor via neighbor queries, degree queries and random edge queries  with probability of success at least $2/3$ requires $\Omega(n^{1-1/p})$ total number of queries for any $S_p = \O(n^p)$.
\ET 
\BT \label{thm-large-bound}
For any constant $p \geq 2$, any (randomized) algorithm for approximating $S_p$ to a multiplicative factor via neighbor queries, degree queries and random edge queries with probability of success at least $2/3$ requires $\Omega\left(\frac{n^{p-1/p}}{S_p^{1-1/p}}\right)$ total number of queries for any $S_p=\Omega(n^p)$.  
\ET




\section{Acknowledgements}
This material is based upon work supported by the National Science Foundation Graduate Research Fellowship under Grant No.~CCF-1217423, CCF-1065125, CCF-1420692, and CCF-1122374.
Any opinion, findings, and conclusions or recommendations expressed in this material are those of the authors(s) and do not necessarily reflect the views of the National Science Foundation.
We thank Peter Haas and Samuel Madden for helpful discussions.

\ifnum\icalp=0
\bibliographystyle{alpha}
\else
\bibliographystyle{plain}
\fi

\bibliography{bib}

\appendix
\section{Useful Inequalities}

This section provides standard equalities that we use throughout our paper. These inequalities exist in many variations, but here we only present the formulations which are most convenient for our purposes.

\BT \label{chebyshev} (Chebyshev's Inequality)
For any random variable $X$ and $a > 0$,
\[\P[|X - \E[X]| \geq a] \leq \frac{\Var[X]}{a^2}\]
\ET

\BT \label{markov} (Markov's Inequality)
For any non-negative random variable $X$ and $a > 0$,
\[\P[X \geq  a] \leq \frac{\E[X]}{a}.\]
\ET

\BT \label{chernoff} (Chernoff Bound)
Let $X_1, \cdots, X_n$ be independent Poisson random variables such that $\P[X_i = 1] = p$ for all $i \in [n]$, and let $X = \frac{1}{n}\sum_{i=1}^{n} X_i$.
Then for any $0 < \delta \leq 1$,
\[\P[X > (1+\delta) p] < e^{-\delta^2pn/3}.\]
\ET

\BT \label{jensen} (Jensen's Inequality)
For any real convex function $f$ with $x_1, \cdots, x_n$ in its domain,
\[\sum_{i=1}^n f(x_i) \geq n f\left(\sum_{i=1}^n x_i\right)\]
\ET
\section{Proof of Lemma~\ref{good-when-crude}} \label{upunpr}

\gwc*
\BPF
Let us first consider the first item.
Since $\Var[Y] \leq \E[Y^2]$, we will focus on establishing an upper bound of $\E[Y^2]$.
We compute
\begin{align*}
 \E[Y^2] &= \sum_{v\in V} \frac{\deg(v)}{2m}\left(\frac{2m}{\deg(v)} {\deg(v) \choose p}\right)^2  
 = 2m \sum_{v\in V} \frac{1}{\deg(v)} {\deg(v) \choose p}^2\\
&\leq 2m \sum_{v\in V} {\deg(v) \choose p}^{2-1/p}
\leq 2m \left( \sum_{v\in V} {\deg(v) \choose p} \right)^{2-1/p}
= 2m S_p^{2-1/p},
\end{align*}
where the first inequality holds because $(\deg(v))^p \geq {\deg(v) \choose p}$.
Rearranging the terms, we have the following relationship:
\[\frac{\E[Y^2]}{S_p^{2}}
\leq \frac{2m}{p S_p^{1/p}}.\]

Now let us consider our average $\bar{Y}$.
Since $Y_i$ are identically distributed, we have
\[\Var[\bar{Y}]
=\Var\left[\frac{1}{k}\sum_{i=1}^{k}Y_i\right]
=\frac{1}{k^2}\Var\left[\sum_{i=1}^{k}Y_i\right]
=\frac{1}{k}\Var[Y]\leq\frac{1}{k}\E[Y^2].\]
By Chebyshev's inequality (Theorem \ref{chebyshev}), we have
\[\Pr[|\bar{Y}-\E[\bar{Y}]|\geq\epsilon S_p]
\leq\frac{\Var[\bar{Y}]}{\epsilon^2 S_p^2}
\leq \frac{1}{k}\cdot\frac{2m}{p \epsilon^2 S_p^{1/p}}.\]
In order to achieve the desired value $\bar{Y}$ such that $(1-\epsilon)S_p \leq \bar{Y} \leq (1+\epsilon)S_p$ with error probability $1/3$, it is sufficient to take $6m\,/\,{p \epsilon^2 S_p^{1/p}}$
samples.
Recall the assumption that $\widetilde{S}_p$ satisfying $\frac{1}{2}S_p \leq \widetilde{S}_p \leq 6S_p$.
Thus, the number of required samples to achieve such bound with probability $1/3$ is 
\[k = \frac{36m}{p \epsilon^2 \widetilde{S}_p^{1/p}}.\]
For the second item, we apply Markov's Inequality (Theorem \ref{markov}) to the given condition to obtain
\[\P\left[\overline{Y} \geq \frac{1}{2}\widetilde{S}_p\right] \leq \frac{\E[\overline{Y}]}{\frac{1}{2}\widetilde{S}_p} = \frac{S_p}{\frac{1}{2}\widetilde{S}_p} < \frac{\frac{1}{6}\widetilde{S}_p}{\frac{1}{2}\widetilde{S}_p} = \frac{1}{3},\]
implying the desired success probability.

Lastly, we substitute $\eps < 1/2$ to obtain the relationship between $\bar{Y}$ and $(1-\eps)\tilde{S}_p$,
which establishes the condition for deciding whether the given $\tilde{S}_p$ is much larger than $S_p$, as desired.
\EPF
\section{Proof of Lower Bounds for Undirected Graphs with $S_p > n$} \label{lowerproof}

In this section we provide the proof of lower bounds claimed in Section~\ref{largeboundsec}.
Firstly, to properly describe the adjacency list representation of the input graphs, we introduce the notion of graph representation.
Next, we state a main lemma (Lemma~\ref{detlem}) that establishes the constraints of parameters $n_1, d_1, n_2, d_2$ that allows us to create hard instances.
We then move on to describe our constructions, including both the distribution for applying Yao's principle, and the implementation of the oracle for answering random edge queries.
We prove our main lemma for our construction, and lastly, we give the appropriate parameters that complete the proof of our lower bounds.

\subsection{Graph Representations}

Consider the following \emph{representation} $L$ of an adjacency list for an undirected graph $G$.
Let us say that each vertex $v_i$ has $\deg(v_i)$ ports numbered $1, \ldots, \deg(v)$ attached, where the $j^{\rm th}$ port of vertex $v_i$ is identify as a pair $(i,j)$, which is used as an index for $L$.
$L$ imposes a perfect matching between these ports; namely, $L(i_1,j_1) = (i_2, j_2)$ indicates that ports $(i_1,j_1)$ and $(i_2, j_2)$ are matched to each other, and this implies $L(i_2,j_2) = (i_1, j_1)$ as well.
We use $L$ to define the adjacency list of our graph; that is, if $L(i_1, j_1) = (i_2, j_2)$ then the $j_1^{\rm th}$ neighbor of $v_{i_1}$ is $v_{i_2}$ (and vice versa).
Note that there can be many such representations of $G$, and some perfect matchings between ports may yield graphs parallel edges or self-loops.
Furthermore, each edge $e$ is associated with a unique pair of matched cells.

\subsection{Main Lemma}

Our proof proceeds in two steps.
First, we show the following lemma that applies to certain parameters of graphs.

\BL \label{detlem}
Let $n_1, d_1, n_2, d_2$ be positive parameters satisfying the following properties: $d_1$ and $n_2$ are even, $n_2 \leq d_1 \leq 2d_2$ and $d_1 + 2d_2 < n_1$.
Let $n = n_1+n_2$, and define the following two families of graphs on $n$ vertices:
\begin{itemize}[noitemsep,nolistsep]
\item $\mathcal{F}_1$: all graphs containing $n_1$ vertices of degree $d_1$ and $n_2$ isolated vertices;
\item $\mathcal{F}_2$: all graphs containing $n_1$ vertices of degree $d_1$ and $n_2$ vertices of degree $d_2$.
\end{itemize}
Let $r = \frac{(d_1+d_2) n_2}{d_1 n_1}$ and $q = o(1/r)$.
Then, there exists a distribution $\mathcal{D}$ of representations of graphs from $\mathcal{F}_1 \cup \mathcal{F}_2$ such that
for any deterministic algorithm $\mathcal{A}$ that makes at most $q$ total neighbor queries, degree queries and random edge queries,
on the graph representation randomly drawn from $\mathcal{D}$,
$\mathcal{A}$ cannot correctly identify whether the given representation is of a graph from $\mathcal{F}_1$ or $\mathcal{F}_2$ with probability at least $2/3$.
\EL

By applying Yao's principle, the following corollary is implied.

\BC \label{randlem}
Let $n_1, d_1, n_2, d_2$ be parameters satisfying the properties specified in Lemma~\ref{detlem}.
Let $s_1 = n_1 {d_1 \choose p}$ and $s_2 = n_1 {d_1 \choose p} + n_2 {d_2 \choose p}$.
If $s_1 = \Theta(f(n,p))$ and $s_2 \geq c \cdot s_1$ for some constant $c > 1$, then any (randomized) algorithm for approximating $S_p$ to a multiplicative factor
via neighbor queries, degree queries and random edge queries with probability of success at least $2/3$ requires $\Omega(q)$ queries for $S_p = \Theta(f(n,p))$.
\EC

As a second step, we propose a few sets of parameters for different ranges of $S_p$.
Applying Corollary~\ref{randlem}, this yields lower bounds for the remaining ranges of $S_p$.

\subsection{Our Constructions}

\subsubsection{Construction of $\mathcal{D}$}

We prove this lemma by explicitly constructing the distribution.

\begin{figure}
        \centering
	\includegraphics[width=0.4\textwidth]{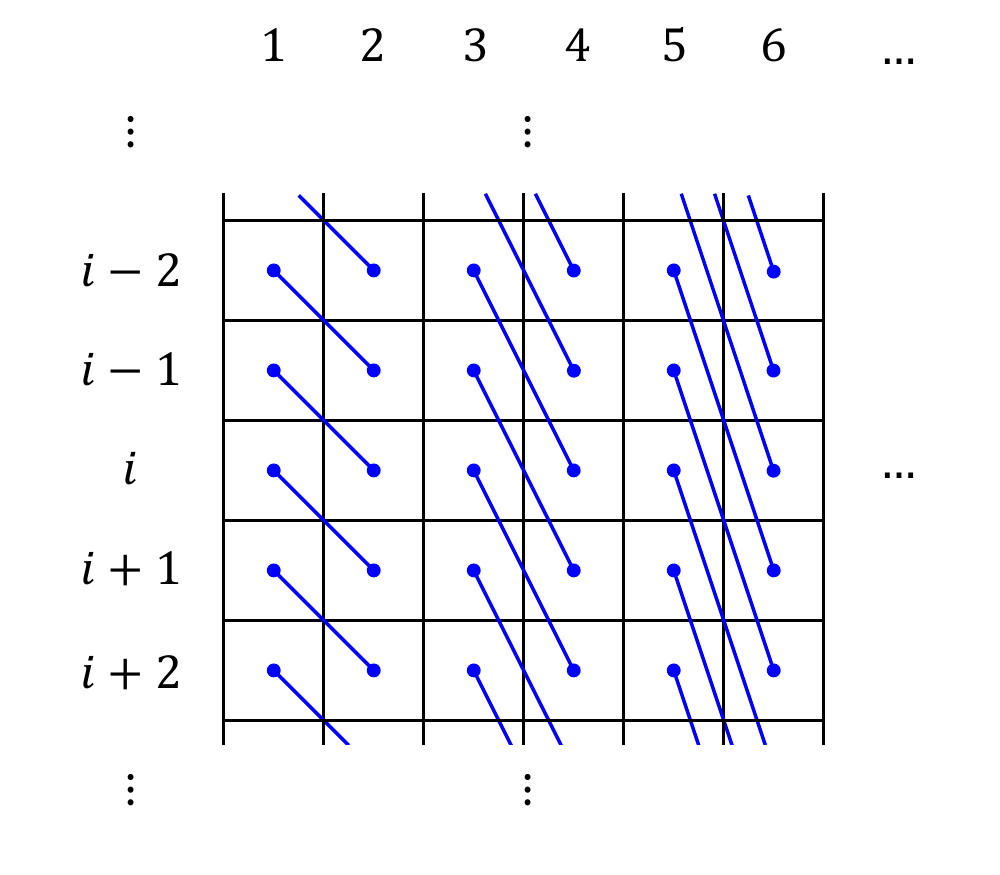}
        \caption{first few columns of $L_1$}
        \label{fig:table1firstcols}
\end{figure}

\noindent\textbf{Construction of graph representations for $\mathcal{F}_1$.}
We now define the representation $L_1$ for the graph $G_1 \in \mathcal{F}_1$ as follows.
We let $v_1, \ldots, v_{n_1}$ be the vertices with degree $d_1$.
Let us refer to the $j^\textrm{th}$ pair of consecutive columns (with indices $2j-1$ and $2j$) as the $j^\textrm{th}$ slab.
Then, in the $j^\textrm{th}$ slab, we match each cell on the left column with the cell at distance $j$ below on the right column.
Figure~\ref{fig:table1firstcols} illustrates the matching of cells in the first few columns of $L_1$.
More formally, for each integer $i \in [n_1]$ and $j \in [d_1/2]$, we match the cells $(i, 2j-1)$ and $(i+j \text{ mod } n_1, 2j)$ in $L_1$.

Since $d_1$ is even, this construction fills the entire table of $L_1$. We wish to claim that we do not create any parallel edges with this construction. Clearly, this is true within a slab. For different slabs,
recall that we map cells in the $j^\textrm{th}$ slab with those at vertical distance $j$ away.
Thus, it suffices to note that no pair of slabs uses the same distance mod $n_1$. Equivalently, we can note that  as the maximum distance is $d_1/2$ and $d_1/2 < n_1/2$ by our assumption, the set of distances $\{j, n_1-j\}$ for $j\in[d_1/2]$ are all disjoint.
That is, our construction creates no parallel edges or self-loops.

\begin{figure}
        \centering
	\includegraphics[width=0.8\textwidth]{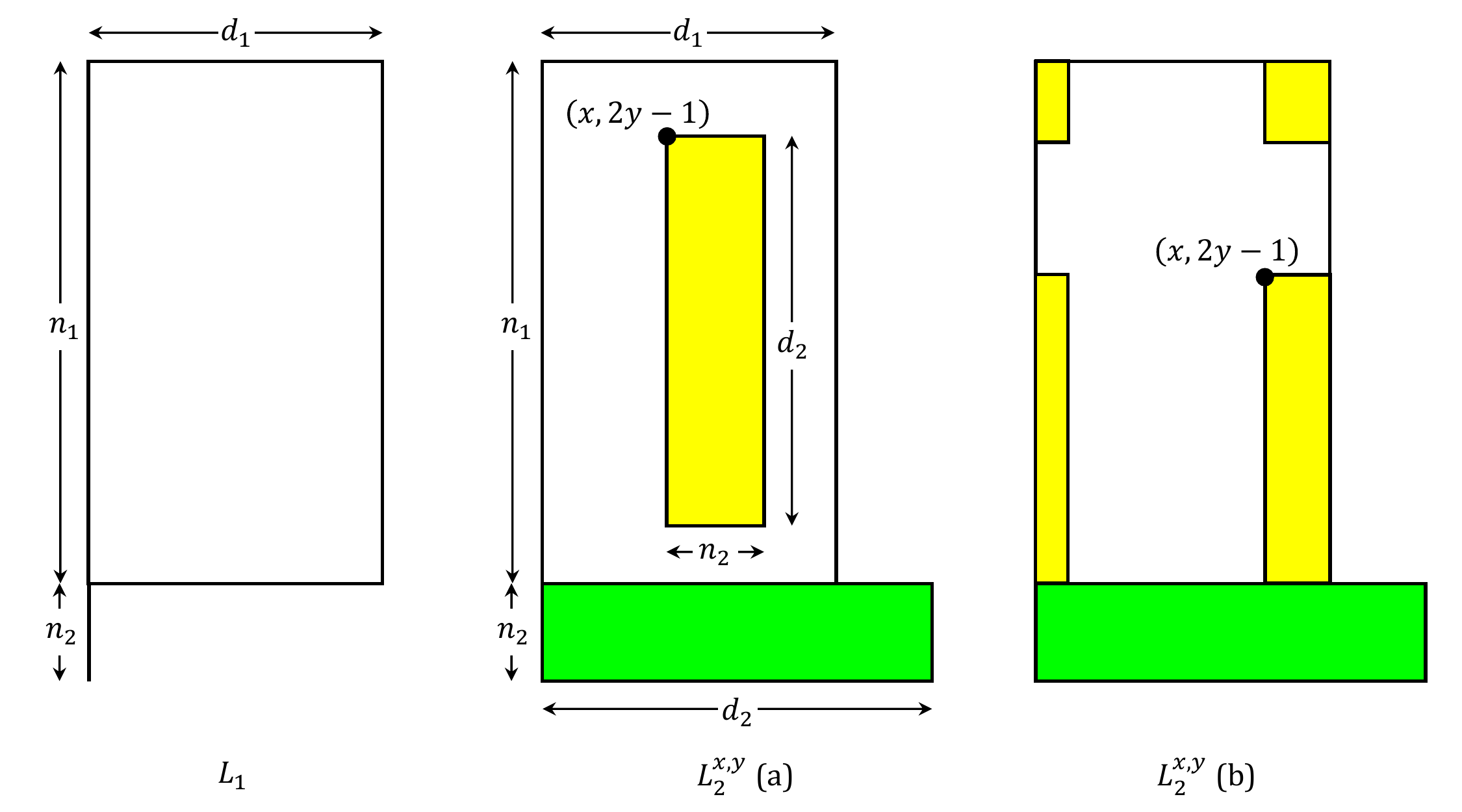}
        \caption{Comparison between tables $L_1$ and $L_2$.  $L_2^{x,y}$ (a) and (b) show two different possibilities for  $L_2^{x,y}$ depending on the values of $x$ and $y$.}
        \label{fig:comparetables}
\end{figure}

\noindent\textbf{Construction of graph representations for $\mathcal{F}_2$.}
Next, for each integer $x \in [n_1]$ and $y \in [d/2]$, we define a graph $G_2^{x, y}$ with corresponding representation $L_2^{x,y}$ by modifying $L_1$ as follows.
First, recall that we need to add neighbors to the previously isolated vertices $v_{n_1+1}, \ldots, v_{n}$.
These neighbors are represented as a table of size $n_2 \times d_2$ in $L_2^{x,y}$;
in Figure~\ref{fig:comparetables}, it is represented as the green rectangle in Figure $L_2^{x,y}$ (a) which is not present in $L_1$.
We match the cells in this new table to a subtable of size $d_2 \times n_2$, which is shown as the yellow rectangle in Figure $L_2^{x,y}$ (a).
The top-left cell of this subtable corresponds to the index $(x, 2y-1)$ in $L_2^{x,y}$,
and note that if $x+d_2 > n_1$ or $2y + n_2 > d_1$, this subtable may wrap around as shown in Figure $L_2^{x,y}$ (b).
Since $n_2 \leq d_1$ and $d_2 < n_1$, the dimensions of this yellow rectangle does not exceed the original table in $L_1$.

\begin{figure}
        \centering
	\includegraphics[width=0.7\textwidth]{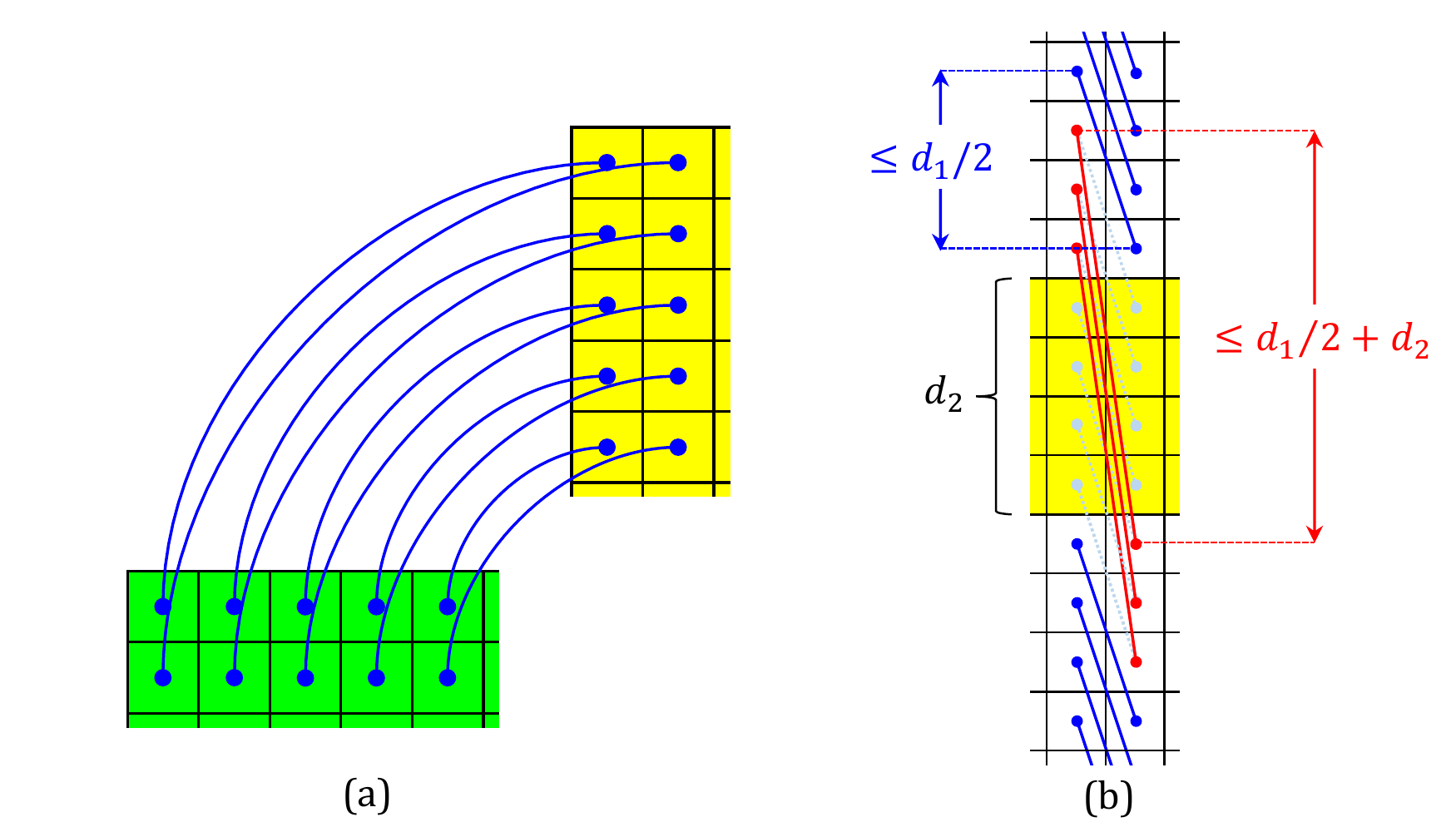}
        \caption{matchings in $L_2^{x,y}$}
        \label{fig:modifiedmatch}
\end{figure}

Now we explain how we match the cells.
Between the yellow and green subtables, we map them in a transposed fashion.
That is, the cell with index $(i, j)$ (relative to the green table) is mapped to the yellow cell with index $(j, i)$ (relative to the yellow subtable),
as shown in Figure~\ref{fig:modifiedmatch} (a).
This method guarantees that no two rows contain two pair of matched cells between them.
As a result, we do not create any parallel edges or self-loops.

As we place the yellow subtable, some edges originally in $L_1$ may now have only one endpoint in the yellow subtable. We refer to the cells in the table that correspond to such edges as \emph{unmatched}.
Since $n_2$ is even and we set our offset to $(x, 2y-1)$, then every slab either does not overlap with the yellow subtable, or overlaps in the exact same rows for both columns of the slab.
Thus, the only edges that have one endpoint in the yellow subtable are those that go from a cell above it to one in it. Roughly speaking, we still map the cells in the same way but ignore the distance it takes to skip over the yellow subtable. More formally, in the $j^\textrm{th}$ slab, we pair each unmatched cell from the left and right respectively that are at vertical distance $j+d_2$ away (instead of $j$), as shown with the red edges in Figure~\ref{fig:modifiedmatch} (b).

Now the set of distances between the cells corresponding to an edge in the $j^\textrm{th}$ slab are $\{j, j+d_2, n_1-j, n_1-(j+d_2)\}$, since distances can be measured both by going down and by going up and looping around.
From our assumption, $d_1/2 \leq d_2$ and $d_1/2 + d_2 < n_1/2$, and thus no distance is shared by multiple slabs, and thus there are no parallel edges or self-loops.

\noindent\textbf{Permutation of graph representations.}
Let $\pi$ be a permutation over $[n]$.\footnote{A permutation $\pi$ over $[n]$ is a bijection $\pi : [n] \rightarrow [n]$.}
Given a graph representation $L$, we define $\pi(L)$ as a new presentation of the same underlying graph, such that the indices of the vertices are permuted according to $\pi$.
We may consider this operation as an interface to the original oracle.
Namely, any query made on a vertex index $i$ is translated into a query for index $\pi(j)$ to the original oracle.
If a vertex index $j$ is an answer from the oracle, then we return $\pi^{-1}(j)$ instead.

\noindent\textbf{The distribution $\mathcal{D}$.}
Let $\mathcal{S}_n$ denote the set of all $n!$ permutations over $[n]$.
We define $\mathcal{D}$ formally as follows:
for any permutation $\pi \in \mathcal{S}_n$, the representation $\pi(L_1)$ corresponding to $G_1$ is drawn from $\mathcal{D}$ with probability $1/(2n!)$,
and each representation $\pi(L_2^{x,y})$ corresponding to $G_2^{x,y}$ is drawn with probability $1/(n_1d_1n!)$ for every $(x,y) \in [n_1]\times[d_1/2]$.
In other words, to draw a random instance from $\mathcal{D}$, we flip an unbiased coin to choose between families $\mathcal{F}_1$ and $\mathcal{F}_2$.
We obtain a representation $L_1$ if we choose $\mathcal{F}_1$; otherwise we pick a random representation $L_2^{x,y}$ for $\mathcal{F}_2$.
Lastly, we apply a random permutation $\pi$ to such representation.

\subsubsection{Answering Random Edge Queries} \label{randedgeoracle}

Notice that Yao's principle allows us to remove randomness used by the algorithm, but the randomness of the oracle remains for the random edge queries.
For any representation we draw from $\mathcal{D}$, the oracle must return an edge uniformly at random for each random edge query.
Nonetheless, we may choose our own implementation of the oracle as long as this condition is ensured.
We apply a coupling argument that imposes dependencies between the behaviors of our oracle between when the underlying graph is from $\mathcal{F}_1$ or $\mathcal{F}_2$.
Let $m_1 = d_1 n_1 /2$ and $m_2 = (d_1 n_1 + d_2 n_2)/2$ denote the number of edges of graphs from $\mathcal{F}_1$ and $\mathcal{F}_2$, respectively.

Our oracle works differently depending on which family the graph comes from.
The following describes the behavior of our oracle for a single query, and note that all queries should be evaluated independently.

\noindent\textbf{Query to $L_1$.}
We simply return an edge chosen uniformly at random.
That is, we pick a random matched pair of cells in $L_1$, and return the vertices corresponding to the rows of those cells.

\noindent\textbf{Query to $L_2^{x,y}$.}
Let $m_s^{x,y}$ denote the number of edges shared by both $L_1$ and $L_2^{x,y}$.
With probability $m_s^{x,y}/m_2$, we return the same edge we choose for $L_1$.
Otherwise, we return an edge chosen uniformly at random from the set of edges in $L_2^{x,y}$ but not in $L_1$.

Our oracle clearly returns an edge chosen uniformly at random from the corresponding representation.
The benefit of using this coupling oracle is that we increase the probability that the same edge is returned to $m_s^{x,y}/m_2$.
By our construction, the cells in $L_1$ that are modified to obtain $L_2^{x,y}$ are fully contained within the subtable of size $(d_1 + d_2) n_2$
obtained by extending the yellow subtable to include $d_1/2$ more rows above and below.
$m_s^{x,y} \geq (d_1 n_1 - (d_1 + d_2) n_2)/2$.
Thus, our oracle may only return a different edge with probability 
\[1-\frac{m_s^{x,y}}{m_2} = 1-\frac{d_1 n_1 - (d_1 + d_2) n_2}{d_1 n_1 + d_2 n_2} = \frac{d_1 n_2}{d_1 n_1 + d_2 n_2} \leq r.\]

\subsection{Proof of Lemma~\ref{detlem}}

Recall that we consider a deterministic algorithm $\mathcal{A}$ that makes at most $q = o(1/r)$ queries.
We may describe the behavior between $\mathcal{A}$ and the oracle with its \emph{query-answer history}.
Notice that since $\mathcal{A}$ is deterministic, if every answer that $\mathcal{A}$ receives from the oracle is the same,
then $\mathcal{A}$ must return the same answer, regardless of the underlying graph.
Our general approach is to show that for most permutations $\pi$, running $\mathcal{A}$ with instance $\pi(L_1)$
will result in the same query-answer history as running with $\pi(L_2^{x,y})$ for most random parameters $\pi$ and $(x,y)$.
If these histories are equivalent, then $\mathcal{A}$ may answer correctly for only roughly half of the distribution.

Throughout this section, we refer to our indices before applying $\pi$ to the representation.
We bound the probability that the query-answer histories are different using an inductive argument as follows.
Suppose that at some point during the execution of $\mathcal{A}$, the history only contains vertices of indices from $[n_1]$,
and all cells in the history are matched in the same way in both $L_1$ and $L_2^{x,y}$.
This inductive hypothesis restricts the possible parameters $\pi$ and $(x,y)$ to those that yield same history up to this point.
We now consider the probability that the next query-answer pair differs, and aim to bound this probability by $O(r)$.

Firstly, we consider a degree query.
By our hypothesis, for a vertex of index outside $[n_1]$ to be queried, $\mathcal{A}$ must specify a vertex it has not chosen before.
Notice that $\mathcal{A}$ may learn about up to 2 vertices from each query-answer pair, so at least $n-2q$ vertices have never appeared in the history.
Since we pick a random permutation $\pi$ for our construction, the probability that the queried vertex has index outside $[n_1]$ is $n_2/(n-2q)$.
As $r \geq n_2/n_1 \geq 1/n_1$, we have $q = o(n_1)$ and our probability simplifies to at most
\[\frac{n_2}{n-2q} = \frac{n_2}{(n_1+n_2)-2\cdot o(n_1)} \leq \frac{n_2}{n_1(1-o(1))} = O(r).\]

Next, we consider a neighbor query.
From the argument above, with probability $1-O(r)$, the queried vertex given by $\mathcal{A}$ has an index from $[n_1]$.
Similarly, $\mathcal{A}$ may learn about up to $2$ cells from each query-answer pair.
Notice that there are  $(d_1 + d_2) n_2$ different possible $(x, y)$  for which each of these cells could be located in the yellow subtable or the two $(d_1/2) \times n_2$ strips above and below it.
As a result, out of $d_1 n_1 - ((d_1 + d_2) n_2) q$ remaining possible locations for the yellow subtable,
the queried cell and the corresponding answer may be in at most $2 (d_1 + d_2) n_2$ of them.
As $(x, y)$ is randomly chosen, the probability that this next query-answer pair is different is at most
\[\frac{2(d_1 + d_2) n_2}{d_1 n_1 - ((d_1 + d_2) n_2) q} = \frac{2r}{1-rq} = \frac{2r}{1-o(1)} = O(r).\]

Lastly, we consider a random edge query.
From the construction in Section~\ref{randedgeoracle} above, the probability that the returned random edge differs is $O(r)$, regardless of the parameters.

From this inductive argument, the probability that the history differs at each step is at most $O(r)$.
As $\mathcal{A}$ only make $q$ queries, the probability that the history differs is at most $q\cdot O(r) = o(1)$.
Thus with probability $1-o(1)$, it is impossible for $\mathcal{A}$ to distinguish whether the underlying graph is from $\mathcal{F}_1$ or $\mathcal{F}_2$.
Since each family is included in $\mathcal{D}$ with probability density $1/2$,
as $\mathcal{A}$ is deterministic, the answer given by $\mathcal{A}$ for these cases is correct for only half of them.
Thus, the probability of $\mathcal{A}$ correctly distinguish between the two graph families is only $1-\frac{1}{2}(1-o(1)) = \frac{1}{2} + o(1)$, as required.

\subsection{Establishing Lower Bounds}



Now we propose the feasible asymptotic parameters according to Lemma~\ref{detlem} and Lemma~\ref{randlem} in order to establish our lower bounds through the following claim.

\BCM
There exists parameters $n_1, d_1, n_2, d_2$ satisfying the properties specified in Lemma~\ref{detlem}, yielding values $s_1, s_2$ satisfying the properties in Lemma~\ref{randlem}, for each of the following cases:
\begin{enumerate}[nolistsep,noitemsep]
\item $n_1 = \Theta(n), d_1 = \Theta((s/n)^{1/p}), n_2 = \Theta(1), d_2 = \Theta(s^{1/p})$ for $f(n,p) =  \O(n^p)$
\item $n_1 = \Theta(n), d_1 = \Theta((s/n)^{1/p}), n_2 = \Theta(s/n^p), d_2 = \Theta(n)$ for $f(n,p) =  \Omega(n^p)$
\end{enumerate}
\ECM

We omit the proof of this claim; our proof only requires straightforward calculation, and a very similar analysis can be found in \cite{gonen2011counting}.
By computing the value $r$ for each case and applying Lemma~\ref{randlem}, we obtain Theorem~\ref{thm-medium-bound} and Theorem~\ref{thm-large-bound}, respectively.
\section{Extension to Directed Graphs} \label{shortdir}

In this section, we extend our model to the directed case. 
Firstly, we formally give the specification of this new model.
Since most of the specification from the undirected graph model given in Section~\ref{prelim} still applies to the directed case, we only explain the differences between these models.
We assume separate adjacency lists for in-neighbors and out-neighbors, allowing for a neighbor about either type of neighbor.
Similarly, a degree query may ask for either of the in-degree or the out-degree.
Random edge queries now return directed edges $(u, v)$; the algorithm knows both the endpoints and the direction.
We focus on the simplest case of stars with mixed directions: approximately counting the number of paths of length two.

Notice the number of stars where all edges point inward or outward can be computed easily by modifying the weighted vertex sampling to sample using in-degree or out-degree respectively
and then applying the algorithm from Section \ref{upper-undirected}.
This works because the numbers of such stars only depend on the in-degrees and the out-degrees, respectively.
Thus, we turn to the problem of counting directed paths of length two as the next simplest case.

\subsection{Lower Bound}
By constructing hard instances similar to those of Lemma~\ref{smallcase}, we obtain a lower bound of $\Omega(n)$.
More formally, letting $L(G)$ denote the number of paths of length two in the directed graph $G$, we prove the following theorem.

\begin{restatable}{thm}{directlow}
\label{directed-bound}
Any (randomized) algorithm for approximating $L(G)$ to a multiplicative factor via neighbor queries, degree queries and random edge queries requires $\Omega(n)$ total number of queries.
In particular, this number of queries is necessary to distinguish the case where $L(G) = 0$ and the case where $L(G) = n$ with probability 2/3.
\end{restatable}

\BPF
Without loss of generality, we assume $n$ is even. Now, we partition the vertex set $V$ into $S$ and $T$ such that $|S| = |T| = n/2$.
Let $\mathcal{G}_1$ be the family of graphs that contains only $G_1$, the complete bipartite graph where every vertex in $S$ has an edge pointing to every vertex in $T$.
Let $\mathcal{G}_2$ be the family of graphs $G_{(t,s)}$ constructed by taking the graph from $\mathcal{G}_1$ and adding one extra back edge $(t, s) \in T \times S$.
Notice that there can be many adjacency list representations of each graph, and this affect the answers to neighbor queries.
We associate each possible adjacency list representation to each graph, and include all possible such representations in the family.

Clearly, $L(G_1) = 0$, whereas $L(G_{(t,s)}) = n$ for every $G_{(t,s)} \in \mathcal{G}_2$.
For any algorithm to distinguish between $\mathcal{G}_1$ and $\mathcal{G}_2$, when given a graph $G_{(t,s)}$ from $\mathcal{G}_2$,
it must be able to detect the vertex $s$ or $t$, the endpoints of the extra edge, with probability at least 2/3.
Otherwise, if neither $s$ nor $t$ is discovered, the subgraph induced by vertices that the algorithm sees from both families would be exactly the same.
The probability of sampling vertices $s$ or $t$ from a vertex sampling, as well as their incident edges from an edge sampling, is $\O(1/n)$.
Similarly, in order to reach $s$ or $t$ from one of their neighbors,
the algorithm must provide the index of $s$ or $t$ in order to make such neighbor query, which may only succeed with probability $\O(1/n)$.
Thus, $\Omega(n)$ samples are required in order to find $s$ or $t$ with probability 2/3, which establishes our lower bound.
\EPF

\subsection{Upper Bound}

For each $v \in V$, define $l(v) = \deg^-(v)\cdot\deg^+(v)$, which represents the number of length two paths whose middle vertex is $v$.
Thus the number of paths of length two, which we aim to approximate, can be written as $L = \sum_{v\in V} l(v)$.
Notice that $2n$ degree queries suffice for exactly computing the number of such paths, already matching the lower bound.
We explore this problem further by making an assumption in attempt to obtain an algorithm that requires $o(n)$ queries.
To this end, we restrict to direct graphs such that there exists a bound on the ratio of in-degree to out-degree.
More specifically, we assume that there exists a value $r \geq 1$ such that $\frac{1}{r} \leq \frac{\deg^-(v)}{\deg^+(v)} \leq r$,
limiting the ratio between the in-degree and the out-degree of any vertex in $G$.

Under this additional assumption, we obtain a sublinear time algorithm by reduction to what is essentially the undirected case.
Our approach is to modify the weighted vertex sampling process via rejection sampling so that
the probability of sampling a vertex $v$ becomes proportional to $\sqrt{l(v)}$,
bringing the sampling probability of each vertex closer to the number of paths centered at that vertex by the rejection sampling method.
Then we use Algorithm \ref{full-algo} to approximate $\sum_{v\in V} (\sqrt{l(v)})^2$, which requires some modification to the algorithm, explained later.
First, we explain the details of our rejection sampling method.

\BCM
In the directed graph model, given weighted vertex sampling and degree queries,
we may generate a random vertex such that each vertex $v$ is returned with probability $\sqrt{l(v)}/\sum_{v'\in V}\sqrt{l(v')}$
by making $\O(r)$ queries in expectation given the aforementioned assumption.
\ECM
\BPF
We draw a random edge sample $(u, v)$, and query for $u$'s in-degree and out-degree.
We return $u$ with probability $\frac{1}{\sqrt{r}}\sqrt{\frac{\deg^-(u)}{\deg^+(u)}}$.
Otherwise, discard $u$ and repeat the process.

Each vertex $u$ is chosen from a random edge sampling is proportional to its in-degree, $\deg^-(u)$.
We only keep $u$ with probability $\frac{1}{\sqrt{r}}\sqrt{\frac{\deg^-(u)}{\deg^+(u)}}$, so the probability that any vertex $u$ is actually returned is proportional to
$\deg^+(u) \cdot \sqrt{\frac{\deg^-(u)}{\deg^+(u)}} = \sqrt{l(u)}$, as desired.
Since $\frac{1}{r} \leq \frac{\deg^-(v)}{\deg^+(v)} \leq r$, we have that $\frac{1}{r} \leq \frac{1}{\sqrt{r}}\sqrt{\frac{\deg^-(v)}{\deg^+(v)}} \leq 1$.
Thus, $\O(r)$ queries are required to generate one such sample.
\EPF

Define $L' = \sum_{v\in V} \sqrt{l(v)}$.
We now have a method to sample a vertex $v$ with probability $\sqrt{l(v)} / L'$ by increasing the time or query complexities only asymptotically by a factor of $r$.
Now we make the following changes to Algorithm \ref{unbiased-algo} so that it approximates $L$.
First, the algorithm should draw random vertices from the new distribution given above.
Second, we redefine $X = \sqrt{l(v)}$ and $Y = X \cdot L' = \sqrt{l(v)} \cdot L'$, so that $\E[Y]=L$.
Note that the value $L'$ can be approximated via essentially the same method as in Section \ref{approx-m}.
The proof of the variance bound (Lemma \ref{good-when-crude}) can be subsequently modified to obtain $\Var[Y] = \O(\sqrt{n}L^2)$.
(This is essentially the problem of approximating the second frequency moment: see \cite{alon1996space} for more details.)
That is, $\O(\sqrt{n})$ samples from this new distribution, or equivalently $\O(r\sqrt{n})$ queries, suffice to obtain a $(1\pm\epsilon)$-approximation of $L$.
This concludes the proof of the following theorem.
\BT \label{directedup}
Assuming there exists some value $r$ such that $\frac{1}{r} \leq \frac{\deg^-(v)}{\deg^+(v)} \leq r$ for every $v \in V$ in the directed graph $G$,
then there exists an algorithm that,
using degree queries and random edge queries,
computes a $(1\pm\epsilon)$-approximation of the number of paths of length two in $G$ with success probability 2/3
using $\O(r\sqrt{n})$ queries.
\ET
\BC
Assuming that the ratio between the in-degree and the out-degree of every vertex in the directed graph $G$ is bounded above and below by a constant, then there exists an algorithm that,
using degree queries and random edge queries,
computes a $(1\pm\epsilon)$-approximation of the number of paths of length two in $G$ with success probability 2/3
using $\O(\sqrt{n})$ queries.
\EC

\end{document}